%% file: neurips_2018.tex
\documentclass{article}

     \usepackage[nonatbib, final]{neurips_2018}

\usepackage[utf8]{inputenc} 
\usepackage[T1]{fontenc}    
\usepackage{hyperref}       
\usepackage{url}            
\usepackage{booktabs}       
\usepackage{amsfonts}       
\usepackage{nicefrac}       
\usepackage{microtype}      

\usepackage{wrapfig}
\usepackage{subcaption}
\usepackage{color}
\usepackage{pgfplots}
\graphicspath{{figures/}}
\usepackage{amsmath}
\newcommand{\mat}[1]{\mathrm{#1}}
\newcommand{\tensor}[1]{\mathrm{\mathbf{#1}}}
\renewcommand{\vec}[1]{\mathbf{#1}}

\title{Towards Fast Biomechanical Modeling of Soft Tissue Using Neural Networks}

\author{
  Felix~Meister
 \thanks{Pattern~Recognition~Lab, Friedrich-Alexander-University~Erlangen-N\"urnberg, \texttt{felix.meister@fau.de}} \\
  Siemens Healthineers\\
  Digital Technology \& Innovation\\
  \texttt{meister.felix.ext@siemens.com} \\
  \And
  Tiziano~Passerini \\
  Siemens Healthineers\\
  Digital Technology \& Innovation\\
  \texttt{tiziano.passerini@siemens.com} \\
  \And
  Viorel~Mihalef \\
  Siemens Healthineers\\
  Digital Technology \& Innovation\\
  \texttt{viorel.mihalef@siemens.com} \\
  \And
  Ahmet~Tuysuzoglu \\
  Siemens Healthineers\\
  Digital Technology \& Innovation\\
  \texttt{ahmet.tuysuzoglu@siemens.com} \\
  \And
  Andreas~Maier \\
  Pattern Recognition Lab\\
  University of Erlangen-Nuremberg\\
  \texttt{andreas.maier@fau.de} \\
  \And
  Tommaso~Mansi\\
  Siemens Healthineers\\
  Digital Technology \& Innovation\\
  \texttt{tommaso.mansi@siemens.com} \\
}

\begin{document}

\maketitle

\begin{abstract}
To date, the simulation of organ deformations for applications like therapy planning or image-guided interventions is calculated by solving the elastodynamics equations. While efficient solvers have been proposed for fast simulations, methods that are both real-time and accurate are still an open challenge. An ideal, interactive solver would be able to provide physically and numerically accurate results at high frame rate, which requires efficient force computation and time integration. Towards this goal, we explore in this paper for the first time the use of neural networks to directly learn the underlying biomechanics. Given a 3D mesh of a soft tissue segmented from medical images, we train a neural network to predict vertex-wise accelerations for a large time step based on the current state of the system. The model is trained using the deformation of a bar under torsion, and evaluated on different motions, geometries, and hyperelastic material models. For predictions of ten times the original time step we observed a mean error of 0.017\,mm $\pm$ 0.014  (0.032) at a mesh size of 50\,mm x 50\,mm x 100\,mm. Predictions at 20dt yield an error of 2.10\,mm $\pm$ 1.73 (4.37) and by further increasing the prediction time step the maximum error rises to 38.3\,mm due to an artificial stiffening. In all experiments our proposed method stayed stable, while the reference solver fails to converge. Our experiments suggest that it is possible to directly learn the mechanical simulation and open further investigations for the direct application of machine learning to speed-up biophysics solvers.
\end{abstract}

\section{Introduction}
Computational modeling of organ biomechanics has been investigated for disease understanding, therapy planning and image-guided interventions \cite{Anderson07:VVS,Delingette06:CMF,Kayvanpour15:TPC,Wittek16:FFE}. For instance, an image-guided surgery system for the resection of hepatocarcinoma may use the physics-based computation of the deformation field from a personalized model of a patient's liver and a sparse set of tracked points to robustly register acquisitions from pre- and intraoperative imaging modalities \cite{Rucker14:MBN}. Simulations rely on numerical solvers that use the time and space discretization of the equations and constitutive laws to compute the deformation. Accurately calculating the motion in real-time is still an open challenge, but necessary for efficient personalization of models which commonly requires iterative approaches. While simple methods like mass-spring systems proved to be fast, the result may not be accurate enough to capture the nonlinear behaviour of organ tissue under large deformations. In contrast, finite element methods (FEM) may yield precise results, but at the price of not being real-time \cite{Meier05:RTD}. Common approaches in the literature to improve the performance comprise the usage of reduced order models or warping techniques \cite{Choi05:MWR,Yang15:EPF}. Ideally, an interactive solver would be able to provide physically and numerically accurate results at large time-steps, with efficient force computations. 

Towards this goal, we explore in this work the application of deep learning to speed-up biomechanical simulations. We hypothesize that a neural network, as a universal function approximator, is able to learn the underlying biomechanics. We evaluate our method by training a neural network on a synthetic dataset with fixed material model and apply it on various geometries, motions, and material models.

\section{Methods}
Computing soft tissue deformation involves solving the underlying dynamics equation $\mat{M}\vec{\ddot{u}} + \mat{D}\vec{\dot{u}} + \mat{K}\vec{u} = \vec{f}_{e}$, where M is the lumped mass matrix, $\mat{D} = \mu \mat{M}$ is the Rayleigh damping, K is the stiffness matrix, and $\mathbf{f_{e}}$ are external forces. $\vec{\ddot{u}}$, $\vec{\dot{u}}$, and $\vec{u}$ are the vertex acceleration, velocity, and displacement, respectively. In this work, the material response is modeled by a standard exponential law derived from \cite{Holzapfel09:CMO}, although any other elastic model can be used. More precisely, the stress-strain energy function writes $\Psi = a/(2b)~\mathrm{exp}(b(I_{1}-3)) + d(J-1)$, where $a$, $b$, and $d$ are free parameters, $I_{1} = \mathrm{tr}(\tensor{C})$  is the first invariant of the deformation tensor $\tensor{C}$, and $J$ the Jacobian determinant of the deformation. Finite element methods yield the solution using either explicit or implicit time integration. Implicit time integration is unconditionally stable, but one must solve a system of linear equations, which is computationally expensive and potentially inaccurate for large time steps. Explicit time integration requires small time steps to be stable and accurate, which results in far more operations to simulate the same interval, but can be easily parallelized and benefit from GPU-based acceleration. Due to these reasons, we consider in this work explicit FEM as the reference and explore the possibility of going beyond their stability limit. 

In this work, the goal is to learn a model that predicts point-wise accelerations $\vec{\ddot{\vec{u}}}_{\Delta t}$ for a large time step $\Delta t$ given the current state of the system $\vec{\Phi}(t)$.  A speed-up is thus achieved if enough computational steps are skipped in between such that it outweighs the network's execution time. At time $t$ we define $\vec{\Phi}(t)$ as the displacement $\vec{u} (t)$, the velocity $\vec{\dot{u}}_{\Delta t} (t)$, and the total force $\vec{f}_T(t)=\vec{f}_e - \mat{K}\vec{u}$, where $\mat{K}\vec{u}$ denotes the internal force. The system is still parameterizable by the set of tissue parameters and boundary conditions by incorporating the total force. The regression task $\mathcal{R}$ is thus formulated as $\mathcal{R}(\vec{\Phi}(t)) = \mathcal{R}( \{\vec{f}_T (t), \vec{\dot{\vec{u}}}_{\Delta t} (t), \vec{u}(t)\} ) = \vec{\ddot{\vec{u}}}_{\Delta t}(t)$. Finally, the displacement is updated as $\vec{u}(t + \Delta t) = \vec{\ddot{\vec{u}}}_{\Delta t}(t) {\Delta t}^2 + 2\vec{u}(t) - \vec{u}({t-\Delta t})$.

\begin{wrapfigure}{r}{0.525\textwidth}
	\centering
	\vspace{-.2cm}
	\resizebox{\linewidth}{!}{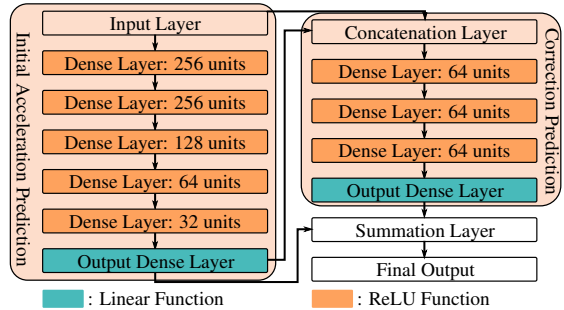} 
	\caption{Architecture details.}
	\label{fig:network}
\end{wrapfigure}
To reduce the number of training samples required for successful learning we further introduce rotation-invariance by transforming the feature vector of each vertex $\vec{v}_i$ at time $t$ into a local coordinate system that is defined by the tangent $\vec{t}_i(t)$, normal $\vec{n}_{i1}(t)$, and bi-normal $\vec{n}_{i2}(t) = \vec{t}_i(t) \times \vec{n}_{i1}(t)$ of its trajectory. For consistency over time we apply the parallel transport algorithm proposed by Hanson and Ma (see \cite{Hanson95:PTA}), which repeatedly rotates an initial coordinate frame at $t =$~0\,s by the angle between two subsequent tangent vectors. By defining the tangent vector as the velocity, we can compress $\vec{\dot{u}}_{\Delta t} (t)$ to its magnitude, leading to a more robust feature vector consisting of seven values, which comprise the 3-D displacement vector, the velocity magnitude, and the 3-D total force vector.

We feed these features point-wise to a network of two functional parts (Figure~\ref{fig:network}). The first one uses five fully-connected layers with a decreasing number of leaky rectified linear units and a linear output layer to make an initial prediction of $\vec{\ddot{\vec{u}}}_{\Delta t}$. To mitigate the error propagation over time, we apply a second network that uses the input and output of the first part as well as three non-linear and a single linear layer to compute the correction. Finally, we sum up both predictions to form the final output, which is clipped to the observed value range to discard outliers. Two separate mean-squared-error loss functions with respect to the ground truth acceleration are applied to both the acceleration prediction and the final output. The network is trained for 200 epochs using Adam  optimizer with a learning rate of 0.001 and exponential decay parameters set to $\beta_1=$~0.9 and $\beta_2=$~0.999 \cite{Kingma14:AMF}.

\section{Experiments \& Discussion}
We generated a synthetic training database by simulating the torsion of a regularly structured rod. The parameters of the exponential law are set to $a=$~0.059, $b=$~8.023, and $c=$~60. The bar consisted of 1458 points and 6528 tetrahedra, with a size of 50\,mm x 50\,mm x 100\,mm. The motion was induced by fixing one side of the bar and applying an external force to the four corner points of the opposite side. The torsion was simulated for one second using the total lagrangian explicit dynamics (TLED) algorithm at a time step of dt~$ = $~3$e^{-5}$\,s, the limit of stability in this configuration \cite{Miller07:TLE}. We discarded the first 0.3\,s due to high-amplitude noise arising from the initial transient state. From the 0.7\,s a total of 1,000 frames were sampled in equal intervals. For each frame we further rejected the points of the fixed plane and the four corner points due to the their discontinuous behaviour caused by the applied boundary conditions. 

\begin{wrapfigure}{L}{0.46\textwidth}
	\centering
	\vspace{-.29cm}
	\resizebox{\linewidth}{!}{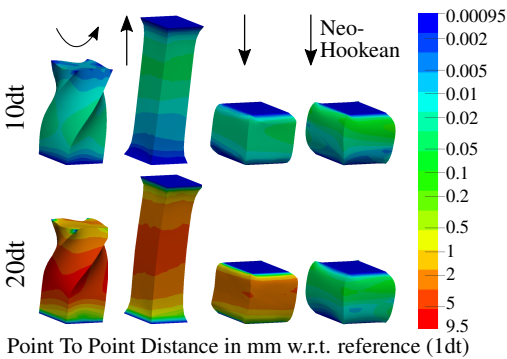} 
	\caption{Error of the computed deformation w.r.t.\ the ground truth for various force induced motions (arrows) calculated with FEM solver at 1dt.}
	\label{fig:proof_of_concept}
\end{wrapfigure}

Our approach was first evaluated by training two networks to predict the acceleration for a time step of 10dt and 20dt, which was hence beyond TLED's stability limit, and applying them to four unseen deformations of the same bar: reversed torsion, stretch, compression, and compression under Neo-Hookean material response. The Neo-Hookean stress strain energy function was defined as $\Psi = c(I_{1}-3) + d(J-1)^2$, whose governing parameters are set to $c =$~0.5 and $d =$~15. To be consistent with the training procedure, we applied a 'burn-in' phase by simulating the first 0.3\,s with TLED at dt before using our neural network at 10dt \& 20dt, respectively. We observed that for a prediction of 10dt the result is almost indistinguishable to the ground truth with a sub-millimeter mean error of 0.017\,mm $\pm$ 0.014  (0.032) over the entire mesh at $t=$~1.0\,s (see Figure~\ref{fig:proof_of_concept}). In contrast, the 20dt prediction exhibits a minor artificial stiffening resulting in a mean error of 2.10\,mm $\pm$ 1.73 (4.37).

\begin{figure}[tb!]
    \centering 
     \resizebox{.9\linewidth}{!}{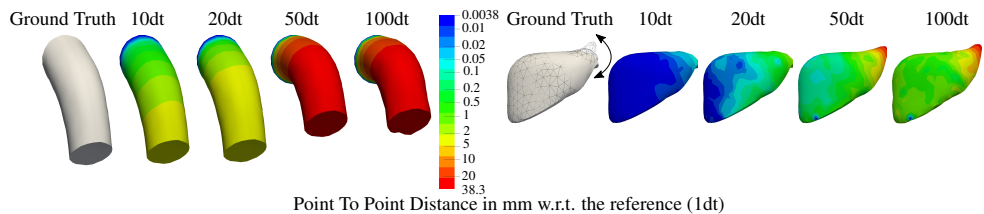} 
     \caption{Illustration of the influence of the time step on the prediction of cylinder bending and liver lobe oscillation experiment. Excellent results up to 20dt, i.e.\ 20 times the inherent time step of the explicit solver, and artificial stiffening for further increased time steps.}
    \label{fig:time_advance}
\end{figure}

Finally, we trained two networks for a time step of 50dt and 100dt on the same rod training set, and applied them to simulate the bending of a cylinder and the oscillation of a liver lobe. The simulations were stable at these time steps, which are far beyond the stability limit of this setup. We discovered that in both cases the artifical stiffening is more profound compared to the 20dt simulation, thus leading to a larger maximum error of 38.3\,mm (see Figure~\ref{fig:time_advance}). This may be due to too few training samples and a too limited capacity of our network to learn the increased complexity.

\section{Conclusion}
In this work we explored the application of deep learning to speed-up the simulation of soft tissue deformation in medical imaging applications. In essence, a neural network is trained to predict point-wise accelerations for a larger time step than possible with traditional explicit finite element methods based on the total force, velocity, and displacement of the current state, which are transformed into a local coordinate system. Our network has been trained on one synthetic simulation and validated under different geometries, motions, and materials. We observed that for a prediction of ten and twenty times the original time step the resulting error is sufficiently small w.r.t.\ the mesh dimension and the proposed method exceeds the stability of the reference FEM solver. While being beyond the stability limit, for larger time advances an artificial stiffening was detected, which is observed in implicit FEM solvers as well. These results suggest that the proposed AI-based solver has both beneficial properties of explicit and implicit FEM methods, namely a straight-forward parallelizibility and an extended stability. To the best of our knowledge, this is the first time a neural network was used to directly learn the biomechanical forces to speed-up the deformation simulations. Possible future research may use recent advances of recurrent neural networks to exploit the sequential nature of the problem for improved accuracy at very high time steps.\\

Disclaimer: This feature is based on research and is not commercially available. Due to regulatory reasons, its future availability cannot be guaranteed.
\small
\bibliographystyle{abbrv}
\bibliography{neurips_2018}

\end{document}

%% file: figures/Network3.pdf_tex
\begingroup%
  \makeatletter%
  \providecommand\color[2][]{%
    \errmessage{(Inkscape) Color is used for the text in Inkscape, but the package 'color.sty' is not loaded}%
    \renewcommand\color[2][]{}%
  }%
  \providecommand\transparent[1]{%
    \errmessage{(Inkscape) Transparency is used (non-zero) for the text in Inkscape, but the package 'transparent.sty' is not loaded}%
    \renewcommand\transparent[1]{}%
  }%
  \providecommand\rotatebox[2]{#2}%
  \ifx\svgwidth\undefined%
    \setlength{\unitlength}{283.38692733bp}%
    \ifx\svgscale\undefined%
      \relax%
    \else%
      \setlength{\unitlength}{\unitlength * \real{\svgscale}}%
    \fi%
  \else%
    \setlength{\unitlength}{\svgwidth}%
  \fi%
  \global\let\svgwidth\undefined%
  \global\let\svgscale\undefined%
  \makeatother%
  \begin{picture}(1,0.55255391)%
    \put(0,0){\includegraphics[width=\unitlength,page=1]{Network3.pdf}}%
    \put(0.01279041,0.50116688){\color[rgb]{0,0,0}\rotatebox{-90}{\makebox(0,0)[lb]{\smash{Initial Acceleration Prediction}}}}%
    \put(0.97188694,0.50851407){\color[rgb]{0,0,0}\rotatebox{-90}{\makebox(0,0)[lb]{\smash{Correction Prediction}}}}%
    \put(0,0){\includegraphics[width=\unitlength,page=2]{Network3.pdf}}%
    \put(0.18149379,0.50410076){\color[rgb]{0,0,0}\makebox(0,0)[lb]{\smash{Input Layer}}}%
    \put(0.09899708,0.43209651){\color[rgb]{0,0,0}\makebox(0,0)[lb]{\smash{Dense Layer: 256 units}}}%
    \put(0.09899708,0.3602866){\color[rgb]{0,0,0}\makebox(0,0)[lb]{\smash{Dense Layer: 256 units}}}%
    \put(0.09899708,0.28845812){\color[rgb]{0,0,0}\makebox(0,0)[lb]{\smash{Dense Layer: 128 units}}}%
    \put(0.10787004,0.21668593){\color[rgb]{0,0,0}\makebox(0,0)[lb]{\smash{Dense Layer: 64 units}}}%
    \put(0.10787004,0.14485736){\color[rgb]{0,0,0}\makebox(0,0)[lb]{\smash{Dense Layer: 32 units}}}%
    \put(0.12271727,0.07304758){\color[rgb]{0,0,0}\makebox(0,0)[lb]{\smash{Output Dense Layer}}}%
    \put(0.60718385,0.48828577){\color[rgb]{0,0,0}\makebox(0,0)[lb]{\smash{Concatenation Layer}}}%
    \put(0.59595657,0.41646978){\color[rgb]{0,0,0}\makebox(0,0)[lb]{\smash{Dense Layer: 64 units}}}%
    \put(0.59595657,0.34465991){\color[rgb]{0,0,0}\makebox(0,0)[lb]{\smash{Dense Layer: 64 units}}}%
    \put(0.59595657,0.27285021){\color[rgb]{0,0,0}\makebox(0,0)[lb]{\smash{Dense Layer: 64 units}}}%
    \put(0.61080381,0.20102152){\color[rgb]{0,0,0}\makebox(0,0)[lb]{\smash{Output Dense Layer}}}%
    \put(0.62689622,0.12923692){\color[rgb]{0,0,0}\makebox(0,0)[lb]{\smash{Summation Layer}}}%
    \put(0.66455866,0.05717619){\color[rgb]{0,0,0}\makebox(0,0)[lb]{\smash{Final Output}}}%
    \put(0,0){\includegraphics[width=\unitlength,page=3]{Network3.pdf}}%
    \put(0.14182227,0.00568957){\color[rgb]{0,0,0}\makebox(0,0)[lb]{\smash{: Linear Function}}}%
    \put(0,0){\includegraphics[width=\unitlength,page=4]{Network3.pdf}}%
    \put(0.62994073,0.00568957){\color[rgb]{0,0,0}\makebox(0,0)[lb]{\smash{: ReLU Function}}}%
  \end{picture}%
\endgroup%

%% file: figures/Motion_symbol_scale_smooth.pdf_tex
\begingroup%
  \makeatletter%
  \providecommand\color[2][]{%
    \errmessage{(Inkscape) Color is used for the text in Inkscape, but the package 'color.sty' is not loaded}%
    \renewcommand\color[2][]{}%
  }%
  \providecommand\transparent[1]{%
    \errmessage{(Inkscape) Transparency is used (non-zero) for the text in Inkscape, but the package 'transparent.sty' is not loaded}%
    \renewcommand\transparent[1]{}%
  }%
  \providecommand\rotatebox[2]{#2}%
  \ifx\svgwidth\undefined%
    \setlength{\unitlength}{373.97110586bp}%
    \ifx\svgscale\undefined%
      \relax%
    \else%
      \setlength{\unitlength}{\unitlength * \real{\svgscale}}%
    \fi%
  \else%
    \setlength{\unitlength}{\svgwidth}%
  \fi%
  \global\let\svgwidth\undefined%
  \global\let\svgscale\undefined%
  \makeatother%
  \begin{picture}(1,0.71102895)%
    \put(0,0){\includegraphics[width=\unitlength,page=1]{Motion_symbol_scale_smooth.pdf}}%
    \put(0.64643657,0.66412343){\color[rgb]{0,0,0}\makebox(0,0)[lb]{\smash{\LARGE Neo-}}}%
    \put(0.90706726,0.6907128){\color[rgb]{0,0,0}\makebox(0,0)[lb]{\smash{\Large 0.00095}}}%
    \put(0.90706726,0.64186422){\color[rgb]{0,0,0}\makebox(0,0)[lb]{\smash{\Large 0.002}}}%
    \put(0.90706726,0.04844546){\color[rgb]{0,0,0}\makebox(0,0)[lb]{\smash{\Large 9.5}}}%
    \put(0.90706726,0.09681814){\color[rgb]{0,0,0}\makebox(0,0)[lb]{\smash{\Large 5}}}%
    \put(0.90706726,0.15603877){\color[rgb]{0,0,0}\makebox(0,0)[lb]{\smash{\Large 2}}}%
    \put(0.90706726,0.2044206){\color[rgb]{0,0,0}\makebox(0,0)[lb]{\smash{\Large 1}}}%
    \put(0.90706726,0.2540228){\color[rgb]{0,0,0}\makebox(0,0)[lb]{\smash{\Large 0.5}}}%
    \put(0.90706726,0.31752185){\color[rgb]{0,0,0}\makebox(0,0)[lb]{\smash{\Large 0.2}}}%
    \put(0.90706726,0.36590363){\color[rgb]{0,0,0}\makebox(0,0)[lb]{\smash{\Large 0.1}}}%
    \put(0.90706726,0.41426879){\color[rgb]{0,0,0}\makebox(0,0)[lb]{\smash{\Large 0.05}}}%
    \put(0.90706726,0.47969312){\color[rgb]{0,0,0}\makebox(0,0)[lb]{\smash{\Large 0.02}}}%
    \put(0.90706726,0.52807475){\color[rgb]{0,0,0}\makebox(0,0)[lb]{\smash{\Large 0.01}}}%
    \put(0.90706726,0.57601203){\color[rgb]{0,0,0}\makebox(0,0)[lb]{\smash{\Large 0.005}}}%
    \put(-0.0009139,0.00537941){\color[rgb]{0,0,0}\makebox(0,0)[lb]{\smash{\LARGE Point To Point Distance in mm w.r.t. reference (1dt)}}}%
    \put(0.03372089,0.4630917){\color[rgb]{0,0,0}\rotatebox{90}{\makebox(0,0)[lb]{\smash{\huge 10dt}}}}%
    \put(0.03448508,0.14003551){\color[rgb]{0,0,0}\rotatebox{90}{\makebox(0,0)[lb]{\smash{\huge 20dt}}}}%
    \put(0.64643657,0.61870065){\color[rgb]{0,0,0}\makebox(0,0)[lb]{\smash{\LARGE Hookean}}}%
  \end{picture}%
\endgroup%

%% file: figures/Combined_smooth.pdf_tex
\begingroup%
  \makeatletter%
  \providecommand\color[2][]{%
    \errmessage{(Inkscape) Color is used for the text in Inkscape, but the package 'color.sty' is not loaded}%
    \renewcommand\color[2][]{}%
  }%
  \providecommand\transparent[1]{%
    \errmessage{(Inkscape) Transparency is used (non-zero) for the text in Inkscape, but the package 'transparent.sty' is not loaded}%
    \renewcommand\transparent[1]{}%
  }%
  \providecommand\rotatebox[2]{#2}%
  \ifx\svgwidth\undefined%
    \setlength{\unitlength}{699.97564361bp}%
    \ifx\svgscale\undefined%
      \relax%
    \else%
      \setlength{\unitlength}{\unitlength * \real{\svgscale}}%
    \fi%
  \else%
    \setlength{\unitlength}{\svgwidth}%
  \fi%
  \global\let\svgwidth\undefined%
  \global\let\svgscale\undefined%
  \makeatother%
  \begin{picture}(1,0.20942107)%
    \put(0.27211235,0.0005882){\color[rgb]{0,0,0}\makebox(0,0)[lb]{\smash{\Large Point To Point Distance in mm w.r.t. the reference (1dt)}}}%
    \put(-0.02791782,0.19683554){\color[rgb]{0,0,0}\makebox(0,0)[lb]{\smash{\Large Ground Truth}}}%
    \put(0.09939831,0.19683554){\color[rgb]{0,0,0}\makebox(0,0)[lb]{\smash{\Large 10dt}}}%
    \put(0.18176235,0.19683554){\color[rgb]{0,0,0}\makebox(0,0)[lb]{\smash{\Large 20dt}}}%
    \put(0.26301828,0.19683554){\color[rgb]{0,0,0}\makebox(0,0)[lb]{\smash{\Large 50dt}}}%
    \put(0.34771452,0.19683554){\color[rgb]{0,0,0}\makebox(0,0)[lb]{\smash{\Large 100dt}}}%
    \put(0.45049344,0.20256969){\color[rgb]{0,0,0}\makebox(0,0)[lb]{\smash{0.0038}}}%
    \put(0.45039492,0.18896929){\color[rgb]{0,0,0}\makebox(0,0)[lb]{\smash{0.01}}}%
    \put(0.45030028,0.17120774){\color[rgb]{0,0,0}\makebox(0,0)[lb]{\smash{0.02}}}%
    \put(0.4503134,0.15669867){\color[rgb]{0,0,0}\makebox(0,0)[lb]{\smash{0.05}}}%
    \put(0.45185028,0.14342665){\color[rgb]{0,0,0}\makebox(0,0)[lb]{\smash{0.1}}}%
    \put(0.45175565,0.12566519){\color[rgb]{0,0,0}\makebox(0,0)[lb]{\smash{0.2}}}%
    \put(0.45176881,0.11062761){\color[rgb]{0,0,0}\makebox(0,0)[lb]{\smash{0.5}}}%
    \put(0.45169319,0.09696146){\color[rgb]{0,0,0}\makebox(0,0)[lb]{\smash{1}}}%
    \put(0.45172877,0.08071197){\color[rgb]{0,0,0}\makebox(0,0)[lb]{\smash{2}}}%
    \put(0.45171314,0.06687123){\color[rgb]{0,0,0}\makebox(0,0)[lb]{\smash{5}}}%
    \put(0.45014486,0.0509616){\color[rgb]{0,0,0}\makebox(0,0)[lb]{\smash{10}}}%
    \put(0.45027576,0.0326449){\color[rgb]{0,0,0}\makebox(0,0)[lb]{\smash{20}}}%
    \put(0.45026235,0.01767653){\color[rgb]{0,0,0}\makebox(0,0)[lb]{\smash{38.3}}}%
    \put(0.49782807,0.1966036){\color[rgb]{0,0,0}\makebox(0,0)[lb]{\smash{\Large Ground Truth}}}%
    \put(0.6365732,0.1966036){\color[rgb]{0,0,0}\makebox(0,0)[lb]{\smash{\Large 10dt}}}%
    \put(0.73493774,0.1966036){\color[rgb]{0,0,0}\makebox(0,0)[lb]{\smash{\Large 20dt}}}%
    \put(0.83905165,0.1966036){\color[rgb]{0,0,0}\makebox(0,0)[lb]{\smash{\Large 50dt}}}%
    \put(0.93974844,0.1966036){\color[rgb]{0,0,0}\makebox(0,0)[lb]{\smash{\Large 100dt}}}%
    \put(0,0){\includegraphics[width=\unitlength,page=1]{Combined_smooth.pdf}}%
  \end{picture}%
\endgroup%